\documentclass[12pt,preprint]{aastex}

\shorttitle{PN Mass Loss in Ellipticals}
\shortauthors{Bregman et al.}

\begin{document}

\title{Mass Loss From Planetary Nebulae in Elliptical Galaxies}

\author{Joel N. Bregman and Joel R. Parriott}

\affil{Department of Astronomy, University of Michigan, Ann Arbor, MI 48109}
\email{jbregman@umich.edu}

\begin{abstract}

Early-type galaxies possess a dilute hot ($2-10\times 10^{6}$ K) gas that is
probably the thermalized ejecta of the mass loss from evolving stars. \ We
investigate the processes by which the mass loss from orbiting stars
interacts with the stationary hot gas for the case of the mass ejected in a
planetary nebula event. \ Numerical hydrodynamic simulations show that at
first, the ejecta expands nearly symmetrically, with an upstream bow shock
in the hot ambient gas. \ At later times, the flow past the ejecta creates
fluid instabilities that cause about half of the ejecta to separate and the
other half to flow more slowly downstream in a narrow wake. \ When radiative
cooling is included, most of the material in the wake ($>80\%$) remains
below $10^{5}$ K while the separated ejecta is hotter ($10^{5}-10^{6}$ K). \
The separated ejecta is still less than one-quarter the temperature of the
ambient medium and the only way it will reach the temperature of the ambient
medium is through turbulent mixing (after the material has left the grid). \
These calculations suggest that a significant fraction of the planetary
nebula ejecta may not become part of the hot ambient material. \ This is in
contrast to our previous calculations for continuous mass loss from giant
stars in which most of the mass loss became hot gas. \ We speculate that
detectable OVI\ emission may be produced, but more sophisticated
calculations will be required to determine the emission spectrum and to
better define the fraction of cooled material.

\end{abstract}


\section{Introduction}

Early-type galaxies can be luminous diffuse X-ray emitters where the
emission is due to hot gas at $2-10 \times 10^6$ K (e.g., \citealt{athey03,mathu03,hump06,dieh07}).  
The origin of the hot gas is modeled
as the normal mass loss from an evolving old population, and possibly a
smaller contribution due to infall of group or cluster gas.  Stellar mass loss
occurs during the evolutionary stages toward the top of the giant and
asymptotic giant branches.  This mass loss only partly defines the
metallicity of the gas, as Type Ia supernovae will contribute at least as
much mass in heavy elements (but an insignificant amount of hydrogen
and helium).  The stellar mass loss occurs at the velocities of the stars and
once this gas is decelerated by its interaction with existing hot gas, it will
have a temperature that is characteristic of the potential well of the galaxy. 
An additional source of heat is provided by supernovae, which may raise
the gas temperature sufficiently so that some or all of the gas is gas flows
out as a galactic wind.  Galactic winds will dramatically lower the volume
emission measure of the galactic gas, which is a leading explanation for
galaxies with low luminosities of diffuse X-ray emission.

An important aspect of this model, the mass loss from stars into the galactic insterstellar medium, needs to be understood as it affects the properties of the hot and cool gas in the galaxy.  The metallicity of the hot medium is expected to result from the mass shed by stars plus SN Ia.  The stellar metallicity of the bright ellipitcals is about solar \citep{trag06}, and if the relative abundance pattern is similar to the Galactic bulge stars\citep{zocc08}, the lighter elements, such as oxygen or magnesium should be enhanced, relative to iron.  However, the elemental abundance measurements in ellipticals, from XMM-Newton and Chandra X-ray data, show that oxygen is typically subsolar \citep{ji09}, less than anticipated by a factor of 2-4 (including the modest contribution of oxygen from SN Ia makes the discrepancy greater).  

Stellar mass loss that does not become part of the hot ambient medium may simply remain neutral or warm ionized (\textless\ 10$^4$ K).  Alternatively, it may become hotter (10$^5$-10$^6$ K), but cools radiatively, never becoming part of the ambient gas at 3-10$\times 10^6$ K \citep{fuji96,mathu90,brig05}.  This could occur because the mass loss from higher
metallicity stars has a greater radiative loss rate, so higher metallicity gas
cools radiatively and does not become part of the hot medium. In both cases, there may be observational consequences as such gas may produce detectable optical or ultraviolet emission lines.

To examine some of these issues, we began a series of numerical hydrodynamic
calculations to study the fate of stellar mass in the hot gas environment of
an early-type galaxy \citep{parr98}.  In our first paper (Paper I, \citealt{parr08}), we discussed the two-dimensional
hydrodynamic code (a version of VH-1), the properties of the initial
conditions, and a set of results for a mass-losing giant star.  In particular,
the mass loss rate was taken to be continuous in time at the rate
$\dot{M}_{star} \approx 10^{-7} M_{\odot} yr^{-1}$, which is representative of stars 
along the giant and asymptotic giant branches.  At this rate, the flow
evolves to a quasi-steady state on a timescale shorter than the mass-loss 
stage of these stars (the duration of our simulations is $3 \times 10^6$ yr).

When the stellar velocity is supersonic relative to the hot gas, a bow shock
developed ahead of the star, imparting momentum to the stellar mass loss,
producing a cool slowly-moving wake behind the star.  There is a large
velocity difference between the wake and the hot ambient gas, so Kelvin-
Helmholtz instabilities develop, which leads to thermal mixing as well as
shocks between these two components.  In the absence of radiative
cooling, the wake is heated to the ambient temperature of the hot medium
within about 2 pc of the star.  When radiative cooling is included, up to
25\% of the wake remains cool by the time it exits the grid, 25 pc from the
star.  The fraction of cooled gas is lower for lower ambient pressures
(densities) or higher stellar velocities, but the fraction of cooled gas was
not tremendously sensitive to the metallicity of the stellar mass loss, at
least for the parameters considered.

A second type of mass loss event occurs in stars, the planetary 
nebula (PN) stage, which is an impulsive mass loss event in the life of a star.  
The planetary nebula mass loss stage typically only lasts $\sim 10^3$ 
years, with mass loss rates near $\sim 10^{-4} M_{\odot} yr^{-1}$, and velocities 
$\sim 10$ km/s \citep{b88,fbr90}.  
Although some details of the mass loss mechanism remain to be clarified, 
the timescale and the amount of mass loss are understood to the degree
necessary for our purposes.  Most studies agree that about $0.2 - 0.3 M_{\odot}$ 
is lost during the red giant and asymptotic giant stage, and another 
$0.1 - 0.2 M_{\odot}$ is lost in the planetary nebula shell (e.g., \citealt{bst94}, \citealt{bsc95}).
In this paper, we simulate the interaction between this planetary nebula 
shell and hot ambient medium through which it is moving.

\section{Simulations}

A detailed discussion of the hydrodynamic code is given in Paper I, so here
we briefly summarize the nature of the calculation and the properties of the
simulations considered.

\subsection{Computational Method and Initial Conditions}

Our two-dimensional hydrodynamic code, VH-1, was originally
developed by the Virginia Institute of Theoretical Astrophysics (VITA) Numerical 
Astrophysics Group \citep{blond93}.  It is based on a version of the 
Piecewise Parabolic Method \citet{cole84}, but with changes to the 
geometry (spherical polar coordinates) and boundary conditions used here.  
Optically thin radiative losses are included through the time-independent 
collisional ionization equilibrium cooling function, $\Lambda_{N}(T)$, of
\citep{sd93}.  This cooling function covers the temperature range
$10^{4} K - 10^{8.5} K$; for cooler gas, radiative losses are set to zero.

A non-uniform grid was utilized in order to have adequate resolution in the 
wake and at radii close to the planetary nebula.  
The outermost radius is 25 pc and at the innermost radius, 0.01 pc, where
the ratio between the radius at \textit{i+1} and \textit{i} is 1.01; there are 360
radial cells.  The azimuthal cell spacing is chosen so that the wake, which
develops along the polar axis, is described by about 100 cells, with the
remainder of the azimuthal cells of nearly equal angular spacing.  
Fairly standard boundary conditions are employed and the code was converted
to run in parallel mode, both topics being discussed in great depth in
Paper I.  A relevant detail is that there is outflow from the inner radial boundary,
in order to permit mass loss.

There is a constant flow entering the grid from the left at a velocity of
$v_{amb} = 350$ km s$^{-1}$, which corresponds to a one-dimensional velocity 
dispersion $\sigma_{\star} \approx 200$ km s$^{-1}$.  This velocity dispersion is 
representative of an intermediate luminosity galaxy.  We set the temperature of the
X-ray gas $T_{amb} = 3 \times 10^{6}\,{\rm K} \approx 0.3\,{\rm keV}$, also 
typical of a $L_{X}$ galaxy.  The density of the ambient gas is taken to be
$n_{amb} = 10^{-3}\,{\rm cm}^{-3}$, which is a value found several kpc from 
the center of an X-ray luminous elliptical and within 1 kpc from the center of
an X-ray poor galaxy.

We simulate the planetary nebula event by a step function increase in the
mass loss rate at the inner radial boundary.  This is accomplished by 
increasing the ejecta density by a factor of $10^3$ for a duration of $5 \times 10^3$ years.
The velocity and temperature input parameters for the ejecta remain 
at $35$ km/s and $10^4$ K, respectively.  This results in an effective
mass loss rate of $\sim 10^{-4} M_{\odot} yr^{-1}$, for a total PN shell mass of 
$\sim 0.5 M_{\odot}$.  We start this simulation using the results of our previous
simulation in which the mass loss rate was constant at the pre-PN rate and run
for $2 \times 10^6$ years (Paper I).
This is an attempt to approximate a more realistic situation where the 
``superwind'' is blown into a previously established wind.  The post-PN mass 
loss rate (i.e. after $5 \times 10^3$ years) reverts back to 
$\sim 10^{-7} M_{\odot} yr^{-1}$.

Since the mass of many PN shells are found to be near $0.1 M_{\odot}$, we
will also carry out an additional set of simulations with a superwind phase
that lasts only $10^3$ years.  The two choices of shell mass bracket the
values observed for PNs.

\subsection{Planetary Nebula Simulation Without Radiative Cooling}
\label{ssec-PNnc}

The evolution scenario for the gas from this simulated PN event begins
with the initial mass loss stage as the thick shell slowly propagates
outward.  The mass loss becomes disrupted by the hot flow past the star 
and is advected downstream and off the grid.  This
entire process takes a physical simulation time of $6 \times 10^5$ years.

In order to accurately know how much of the cold PN gas is heated 
by the time it leaves the grid, we would need to mark the PN material 
and record the temperature of each parcel of gas as it exits the grid, but we did not implement this capability.
Also, since this is a transient event we cannot use the 
time-averaged mass fluxes as we did for all of the previous quasi-steady state
simulation analyses (Paper I).  Rather, in addition to examining individual stages of the 
PN evolution, we will calculate an approximate heating efficiency by recording
the total grid mass, and the mass flux through the outer grid boundary, as a
function of time for both warm gas ($T_{amb}/4$) and cold ($T_{amb}/30$) gas 
temperature bins.

\begin{figure}
\epsscale{0.85}
\plotone{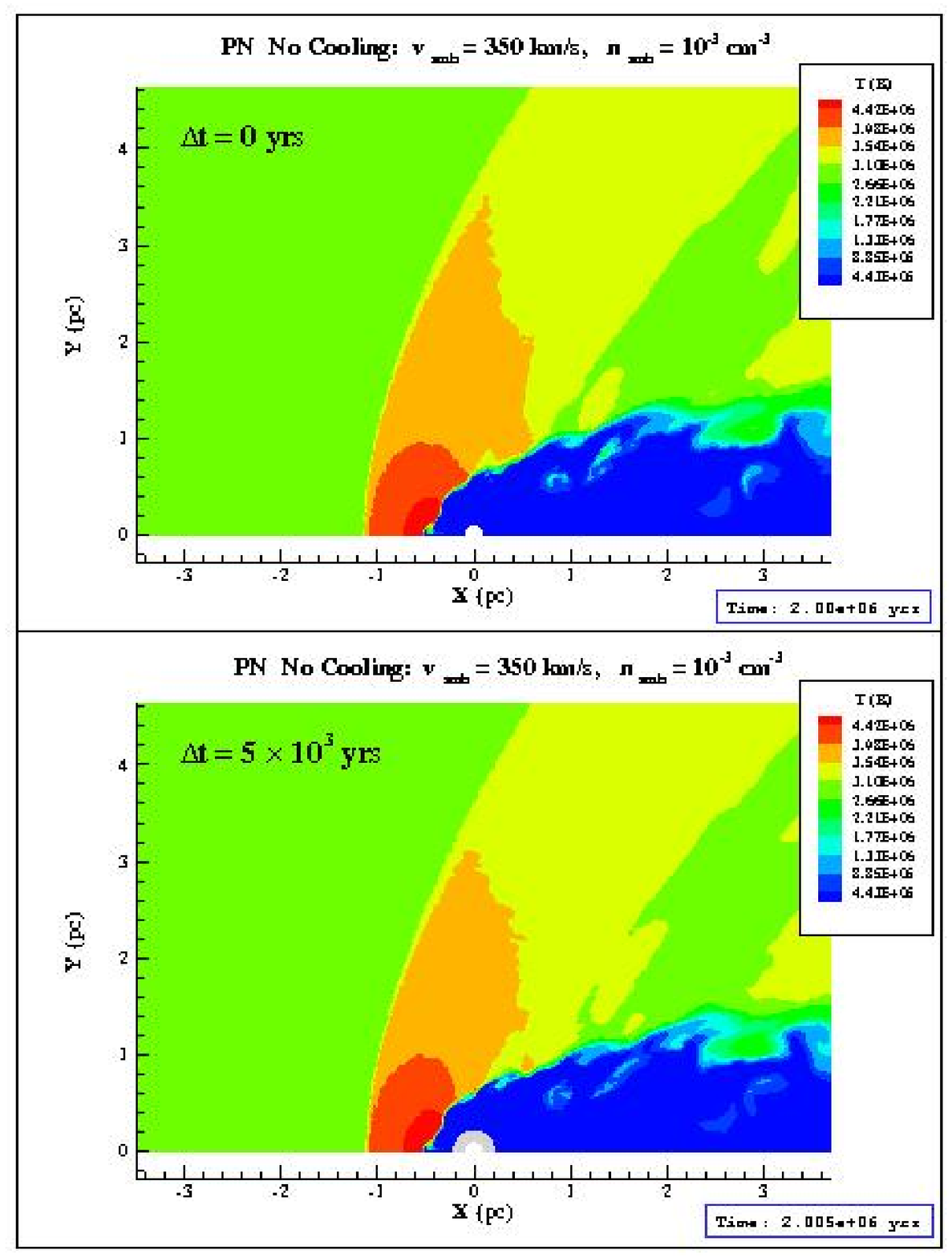}
\caption[The beginning of the PN evolution]%
        {The beginning of the PN evolution just before (top) and just after 
        (bottom) the impulsive superwind phase.  Prior to the PN ejection 
        event, the star was losing mass at a rate of $\sim 10^{-7} M_{\odot} yr^{-1}$,
        typical of a AGB star.  At this point, the cool wake material is from this
        AGB stage.  The velocity difference between ambient hot material and the
        star is 350 km s$^{-1}$, which leads to the bow shock on the upstream side
        of the flow.}
\label{fig:ncpn0.5}
\end{figure}

Beginning with the quasi-steady-state from the continuous mass loss solution
(Figure~\ref{fig:ncpn0.5}), there is a well-established bow shock upstream of the
star.  Momentum transfer from the post-shock hot medium to the mass loss leads
to the slow-moving wake.  The high-velocity flow past the wake produces the cold
extensions upward from the wake due to the excitation of Kelvin-Helmholtz instabilities.
This leads to a combination of mixing and further shocks in the downstream region between the wake and the ambient flow,
ultimately heating the wake and bringing it closer to the velocity of the hot ambient
medium.  Into this flow, the planetary nebula event occurs and is completed by 
$5 \times 10^3$ years (bottom panel of Figure~\ref{fig:ncpn0.5}, where it is a dense
shell very close to the star).

\begin{figure}
\epsscale{0.85}
\plotone{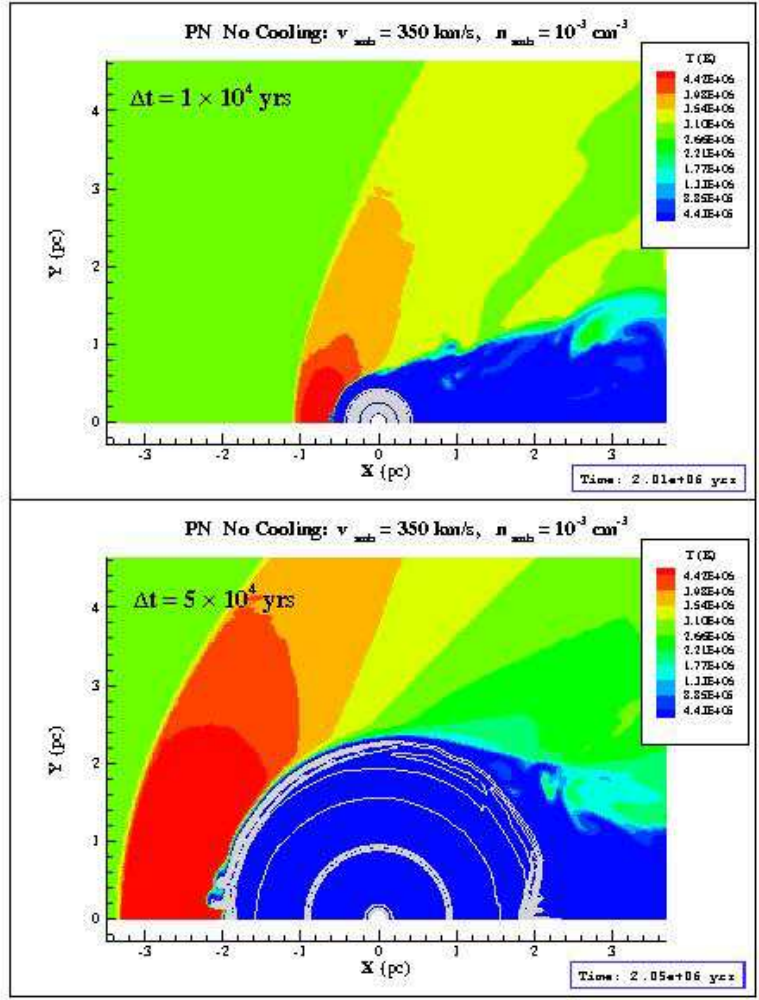}
\caption[Continuing outward flow of the dense PN shell]%
        {Continuing outward flow of the dense PN shell before coming into
        ram pressure equilibrium with the ambient flow.  The times are at
        $10^4$ and $5 \times 10^4$ years after the start of the superwind.}
\label{fig:ncpn10.50}
\end{figure}

This dense shell has enormous momentum compared to its surroundings, so it expands
unimpeded for about $10^5$ years.  During this period, the location of the contact discontinuity moves upstream and the mass flux 
of shocked hot material increases (Figure~\ref{fig:ncpn10.50}).  The
interaction affects both the cool ejecta and the hot ambient medium.
  A reverse shock propagates into the planetary nebula ejecta, while the bow shock heats the dilute ambient material.  These changes continue as the planetary
ejecta expands, and by $5 \times 10^4$ years (bottom panel of Figure~\ref{fig:ncpn10.50}),
the vertical extent of the ejecta has become larger than the original wake height.
This modifies the downstream flow, changing the development of instabilities in the
original wake, although those events are unimportant compared to the evolution of
the much more massive PN ejecta.

\begin{figure}
\epsscale{0.85}
\plotone{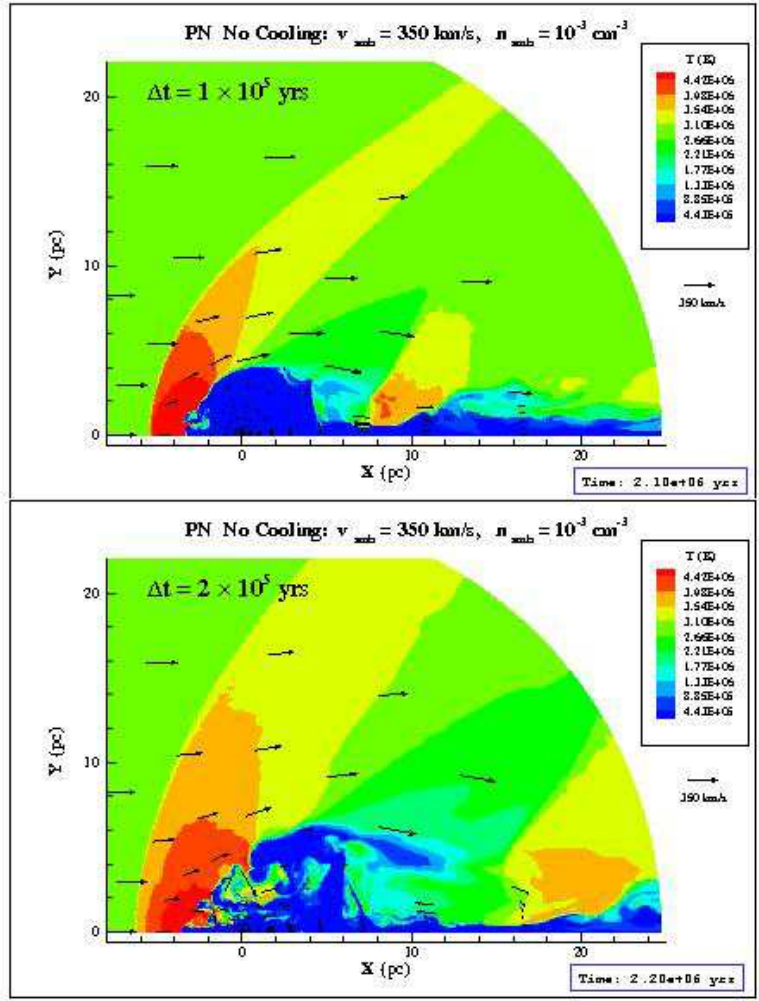}
\caption[Shell expansion ceases and instabilities begin to grow]%
        {Shell expansion ceases and instabilities begin to grow by $10^5$ years
        (top).  The ambient flow has produced instabilities in the
        leading edge of the PN shell.  Disruption of the ejecta is well 
        underway by $2 \times 10^5$ years (bottom).}
\label{fig:ncpn100.200}
\end{figure}

At about $10^5$ years, the shell reaches its maximum radius (about 2.5 pc
in the upstream direction and 4 pc in the vertical direction) as the momentum
density of the shell balances that of the post-shock ambient gas (top panel of
Figure~\ref{fig:ncpn100.200}).  A Rayleigh-Taylor instability has
begun to grow along the leading edge.  The low density post-PN wind 
behind the dense PN shell has a low temperature of order $\sim 10^2$ K, and
a pressure several orders of magnitude below that of the post-shock ambient
gas.  A circular back-flow develops in the dilute ambient gas downstream of
the PN ejecta, which will have a disruptive effect as the ejecta further expands.
By $2 \times 10^5$ years (bottom panel of Figure~\ref{fig:ncpn100.200}), the 
leading edge instabilities have completely broken through the PN shell, 
allowing the ambient gas to enter the bubble and to disrupt the nebula.  
The ambient wind has imparted considerable momentum to the nebula and 
accelerated its velocity to about half the ambient value, so the disruption of 
the PN continues as the entire structure moves downstream at $\sim 150$ km/s,
although the base remains loosely attached to the main downstream wake.

\begin{figure}
\epsscale{0.85}
\plotone{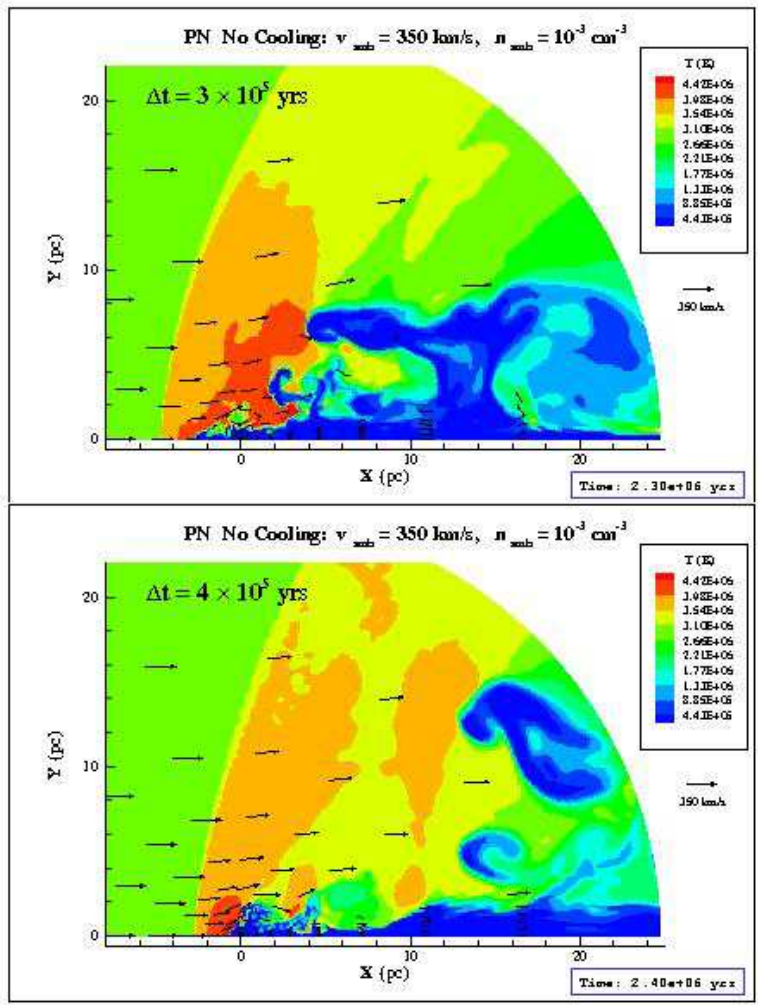}
\caption[The disrupted PN is accelerated downstream and off the grid]%
        {The disrupted PN is accelerated downstream and off the grid by
        momentum transfer from the ambient gas.  Although there is still a
        tenuous connection to the wake, the bulk of the nebula is moving off
        the grid with a velocity of half the ambient value.  The output
        at $3 \times 10^5$ years (top) is the last one with the full nebula
        still on the grid.  By $4 \times 10^5$ years (bottom) one large 
        warm piece of nebula material, with mass $\approx 0.25 M_{\odot}$ or half 
        the original PN mass,
        has separated and the rest of the PN has been advected off the grid.
        There is PN material still in the slower-moving wake.}
\label{fig:ncpn300.400}
\end{figure}

By $3 \times 10^5$ years (Figure~\ref{fig:ncpn300.400}), the base of the 
nebula is roughly half way downstream to the numerical boundary.  The head of the nebula is still 
attached to this base, but ambient gas continues to shock and penetrate
the cloud on the leading edge.  A large vortex has been established on the
back side of the nebula.  This is the last output time where almost all of the
PN gas is still entirely on the grid.  At $4 \times 10^5$ years (bottom panel in
Figure~\ref{fig:ncpn300.400}), the slower moving
left-side structure from the previous figure has been completely separated 
from the rest of the nebula.  This blob has risen more than $10$ pc above the
symmetry axis, and is about to exit the grid.  There has been significant
thermal heating through shocks between the blob and the more rapidly flowing ambient medium as well as through the mixing of the ejecta with the hot ambient gas.  In
this detached cloud, there is no gas cooler than $10^5$ K.
Although some of the original PN material still
resides in the slower moving wake, well over half of the nebula has left
the grid by the next figure at $5 \times 10^5$ years (top panel in 
Figure~\ref{fig:ncpn500.600}).  The PN ejecta has nearly all left the grid by
$5 \times 10^5$ (bottom panel in Figure~\ref{fig:ncpn500.600}) and the flow is 
returning to the quasi-steady properties of the pre-ejecta situation.

In trying to quantify the heating of the PN ejecta, we examined the following 
quantities, taken every $10^4$ years, and for the gas at or below the temperature cutoffs
of $T_{amb}/4 = 7.5 \times 10^5$ K, and $T_{amb}/30 = 10^5$ K:  the 
stellar wind mass loss rate; the total grid mass; and the mass flux through 
the outer grid boundary.
The methods of computing the mass fluxes were described in Paper I.
The total grid mass was computed by summing the product of the mass density 
and the cell volume for each cell with a temperature below the given temperature cutoff.
The background level of wake material present from times before and after 
the PN superwind phase cannot be distinguished from the nebula material, but
this is only a $\sim 4-10\%$ effect since the wake mass is so small compared to
the PN shell mass.

\begin{figure}
\plotone{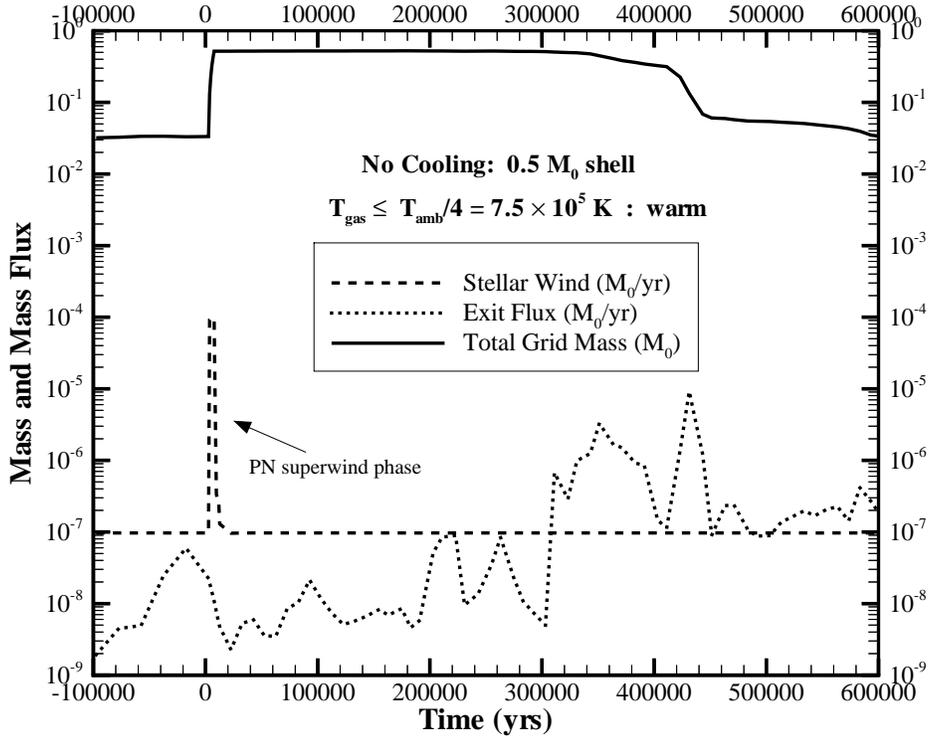}
\caption[Mass fluxes for warm $< T_{amb}/4$ K gas]%
        {Mass fluxes for warm $< T_{amb}/4$ K gas.  The total
        mass and exiting mass flux data suggest that a large fraction of this
        warm PN-material is not heated to ambient temperatures by the time
        it leaves the grid.  The bulk of the nebula begins to exit the grid
        $3 \times 10^5$ years after the superwind.}
\label{fig:nc0.5pnwarmmass}
\end{figure}

The mass flux and total gas mass below a temperature of $T_{amb}/4 = 7.5 \times 
10^5$ K is shown in Figure~\ref{fig:nc0.5pnwarmmass} (we will refer to this as 
warm gas, and as we show, it is dominated by gas in the range $1-7.5\times 10^5$ K).
Due to the temperature threshold, these quantities exclude the ambient gas that 
was heated by the bow shock.  We have also included some simulation data
from $10^5$ years before the PN superwind phase in order to establish the 
background level of wake material.  This figure quantifies the major features of 
the simulations:  there is a time lag of $3 \times 10^5$ years between the
PN superwind phase and the increase of the exiting mass flux above the 
background level (indicating the bulk departure of the PN material).  The
total grid mass data also confirms a background warm gas grid mass level of 
$\sim 0.02 M_{\odot}$, prior to the PN event that releases $0.5 M_{\odot}$.
At least half of the PN ejecta still lies below $7.5 \times 10^5$ K, four times 
cooler than the ambient material.  The large piece of the nebula that leaves the 
grid at $4 \times 10^5$ years has a mass of $\approx 0.25 M_{\odot}$, and additional
warm material leaves the grid near the center of the wake.

\begin{figure}
\plotone{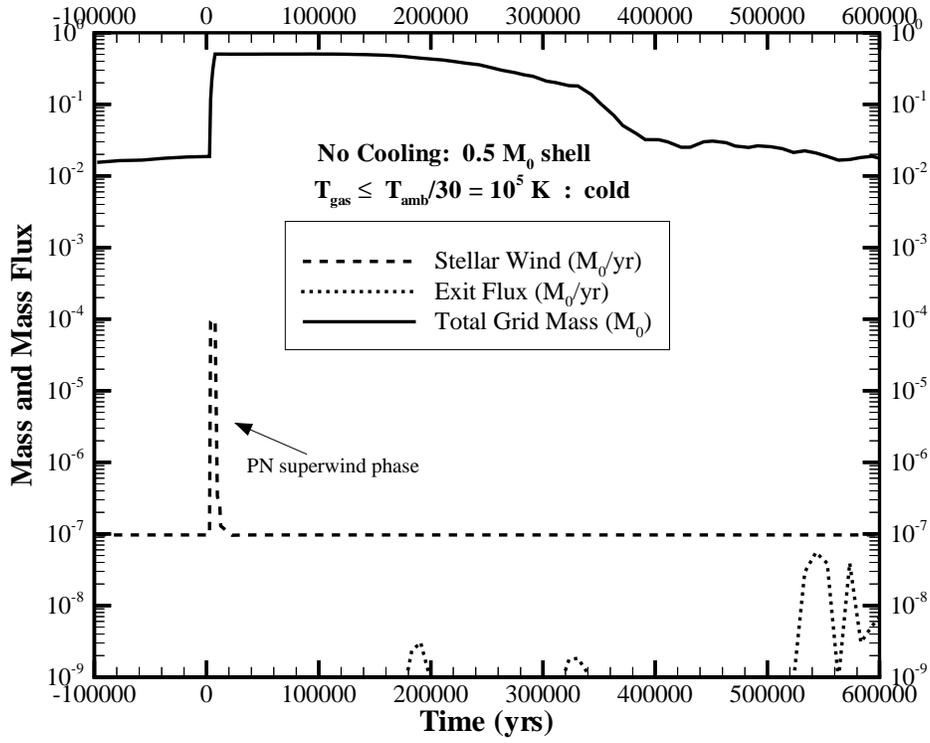}
\caption[Mass fluxes for colder $< T_{amb}/30$ K gas]%
        {Mass fluxes for colder $< T_{amb}/30$ K gas.  These curves
        suggest that this cooler gas is heated by the time it leaves the grid,
        as we would expect for an adiabatic flow.}
\label{fig:nc0.5pncoldmass}
\end{figure}

We also consider the mass of material that is closer to the temperature of the
ejecta ($10^4$ K) by calculating the same quantities for gas at a temperature
below $T_{amb}/30 = 10^5$ K gas (Figure~\ref{fig:nc0.5pncoldmass}; referred to as cold gas).  
The initial shock that is propagated into the PN ejecta heated about three-quarters 
of the gas above $10^5$ K.  Once instabilities became well-developed, at about 
$10^5$ years, the remaining cold gas was quickly heated so that by $2.3 \times 10^5$ 
years, there is no significant amount of cold gas beyond the amount present prior
to the PN event.  To summarize, in the interaction between the PN ejecta and the 
rapidly-moving hot ambient material, the temperature of the ejecta has been raised 
by a factor of 50-100 by the time it exits the grid.  It still is a factor of several below the hot
ambient medium, where the temperature rose due to the bow shock interaction.  
Below, we comment on the likely fate of the warm gas once it leaves the grid.

\subsection{Planetary Nebula Simulation with Cooling}
\label{ssec-PNc}

\begin{figure}
\epsscale{0.85}
\plotone{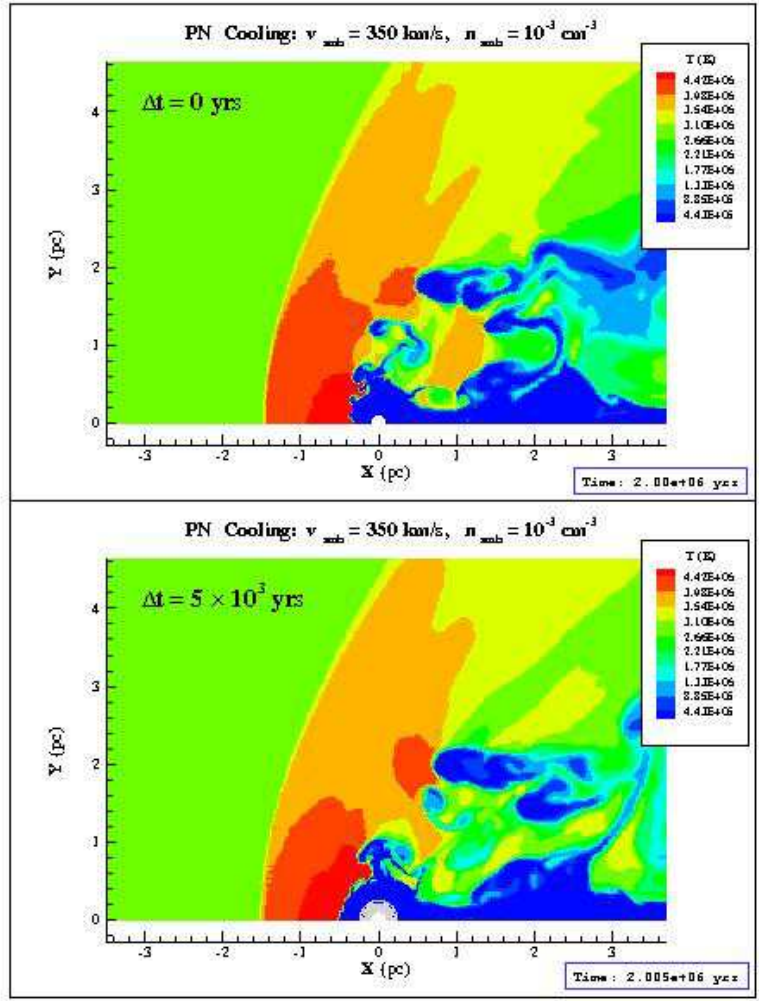}
\caption[The beginning of the PN evolution]%
        {The beginning of the PN evolution just before (top) and just after 
        (bottom) the impulsive superwind phase.  The cooled flow is more
        complex, as discussed with previous simulations.}
\label{fig:cpn0.5}
\end{figure}

In the continuous mass loss case (Paper I), the addition of radiative cooling
changed the structure, as well as the temperature distribution of the mass
shed by the star.  One change was that the Kelvin-Helmholtz and Rayleigh-Taylor
instabilities develop more rapidly and closer to the star, as is seen in
Figure~\ref{fig:cpn0.5}, which forms the initial conditions for this run.
A second change was that the wake became thinner and cooler, with 15\% of
stellar ejecta exiting the grid as cool material (T $< 10^5$ K).  

\begin{figure}
\epsscale{0.85}
\plotone{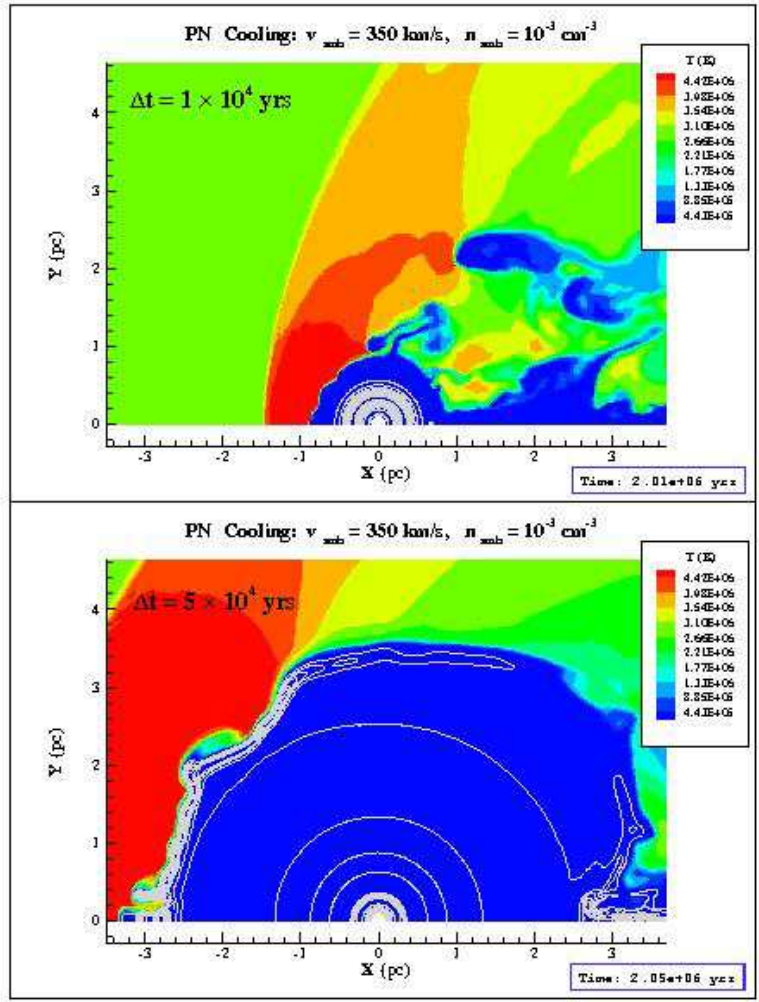}
\caption[Continuing outward flow of the dense PN shell]%
        {The outward flow of the dense PN shell continues before coming into
        ram pressure equilibrium with the ambient flow.  The times are at
        $10^4$ and $5 \times 10^4$ years after the start of the superwind.
        There are more growing leading-edge instabilities in this cooled case.}
\label{fig:cpn10.50}
\end{figure}

Despite the difference in the initial conditions, the dominant feature for
the first $5 \times 10^5$ years is the expansion of the PN ejecta, although the 
timescales (or sizescales) are a bit accelerated for the case with radiative
cooling.  For the radiative cooling case, instabilities develop that 
cause the ejecta event to extend to larger vertical heights and this provides
a larger cross section to the incoming shocked hot flow (Figure~\ref{fig:cpn10.50}).  
In turn, the larger cross section leads to a larger amount of momentum 
transfer, as well as a greater mass of shocked ambient material.

\begin{figure}
\epsscale{0.85}
\plotone{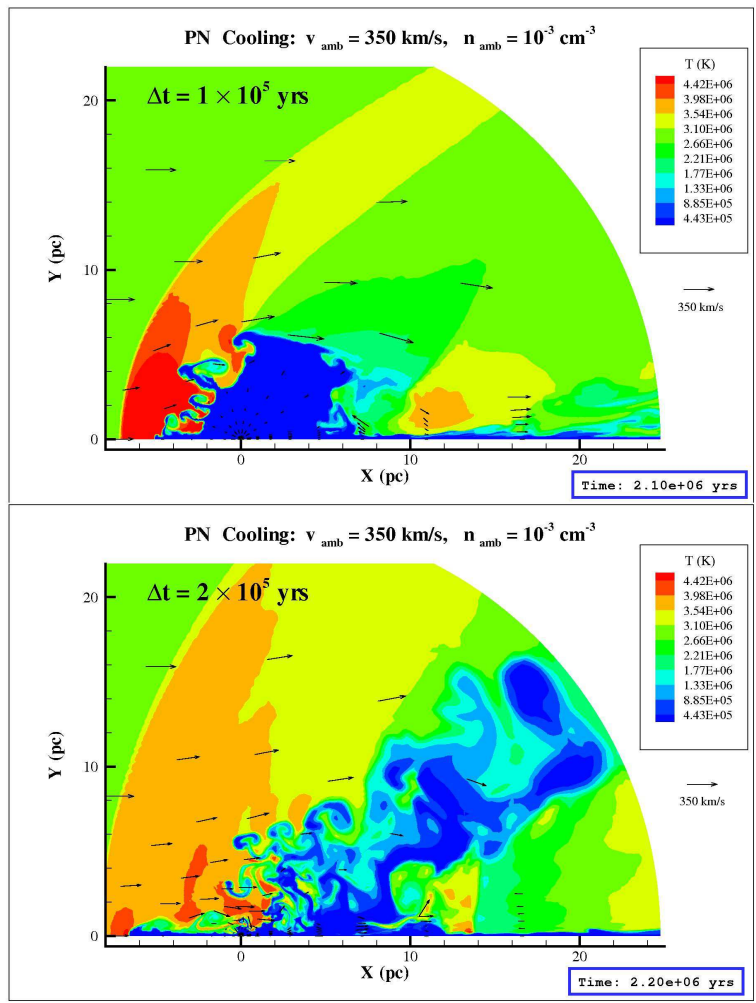}
\caption[Large instabilities have begun nebula disruption]%
        {Large instabilities have begun nebula disruption by $10^5$ years
        (top); the cooled case is much more unstable than the adiabatic case.
        The total disruption of the nebula is achieved by $2 \times 10^5$
        years.  The higher instability of
        the cooled case has made more material available for rapid transfer
        off the grid.  These blobs are more separated from the axial
        wake flow.  The mass of the majority of the disconnected
        material is $0.2 M_{\odot}$, of which $0.02 M_{\odot}$ has a temperature
        below $10^5$ K, so it is colder than the blob in the non-cooling case.
        There is likely to be even more cooled PN material still in the 
        slower-moving wake.  Also, the upstream symmetry axis instability has 
        been amplified by the larger mass outflow.}
\label{fig:cpn100.200}
\end{figure}

\begin{figure}
\epsscale{0.85}
\plotone{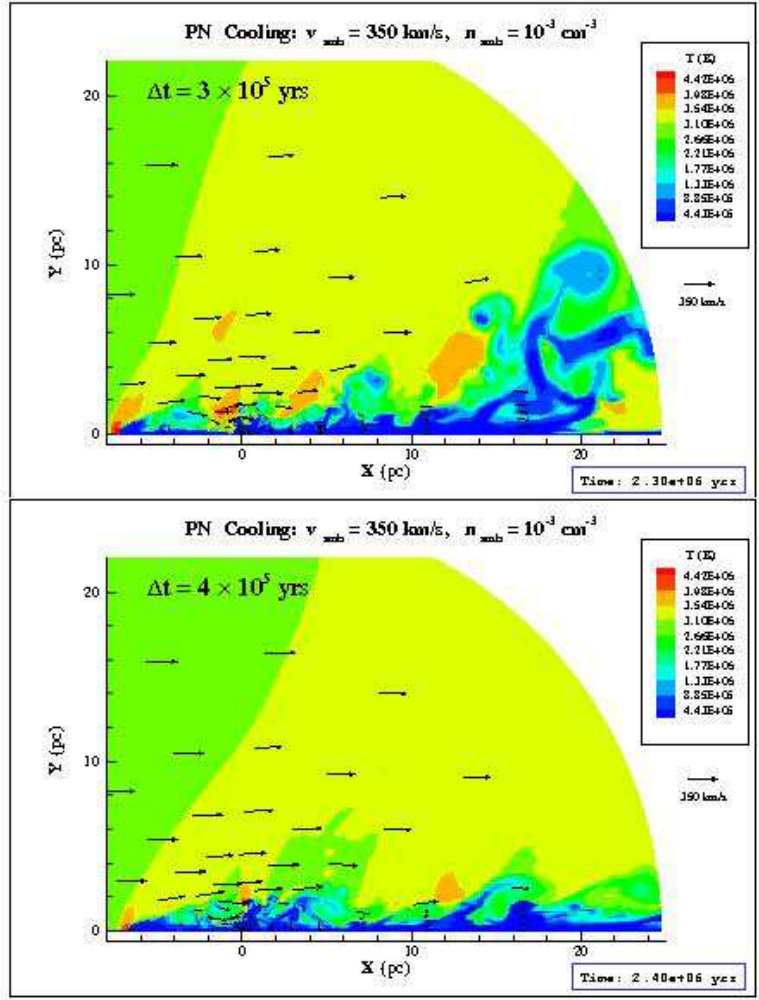}
\caption[The last of the larger cooled structures will exit the grid]%
        {The last of the larger cooled structures will exit the grid soon after
        $3 \times 10^5$ years (top), and the only remaining cooled PN material at
        $4 \times 10^5$ years (bottom) is in the narrow wake region.}
\label{fig:cpn300.400}
\end{figure}

By $10^5$ years, the momentum transfer has pushed the ejecta upward (to about 6 pc)
and downstream, with a number of instabilities developing in the leading surface
of the PN ejecta.  In the subsequent $10^5$ years, the development of fluid
instabilities and momentum transfer have accelerated (bottom panel of 
Figure~\ref{fig:cpn100.200}).  The most striking structures are those that have
nearly entirely broken away from the flow closer to the wake.  These extended
structures lie about 17 pc above the wake as they approach the downstream boundary,
25 pc from the star.  The pieces of PN ejecta that are furthest from the star
have received the most momentum and energy transfer, so one expects them to
have a higher mean temperature than the material in the wake and closer to
the star.  This expectation is realized and when we calculate the mass of 
the of this nebula material that is moving off the grid at about $2 \times 10^5$ years, 
we find a mass of $\approx 0.2 M_{\odot}$ below $7.5 \times 10^5$ K, and 
$\approx 0.02 M_{\odot}$ of that amount is below $10^5$ K.  The last
of the larger warm/cool structures is about to exit the grid at $3 \times 10^5$ years
(Figure~\ref{fig:cpn300.400}), and the bulk of the PN material appears
to be gone by $4 \times 10^5$ years.  While this amount
of cool gas may seem rather small, it is significant in comparison to the 
simulation without radiative cooling, where there was no cool gas in the 
exiting ejecta fragments.  Also, it is possible that radiative cooling will become 
more important in this cool gas as it flows further downstream.

\begin{figure}
\plotone{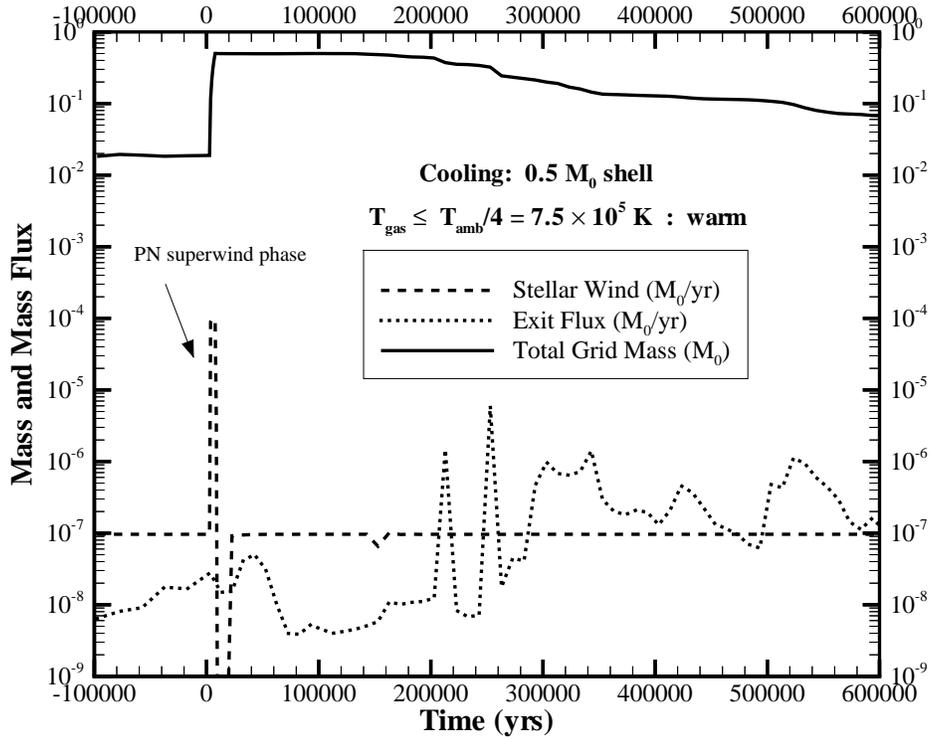}
\caption[Mass fluxes for warm $< T_{amb}/4$ K gas with cooling]%
        {Mass fluxes for warm $< T_{amb}/4$ K gas in the simulations with
        radiative with cooling.  This
        gas is quickly accelerated and advected off the grid by the ambient
        flow, so that the warm gas is gone by $3.5 \times 10^5$ years.  The
        dip after the superwind is a numerical artifact.}
\label{fig:c0.5pnwarmmass}
\end{figure}

\begin{figure}
\plotone{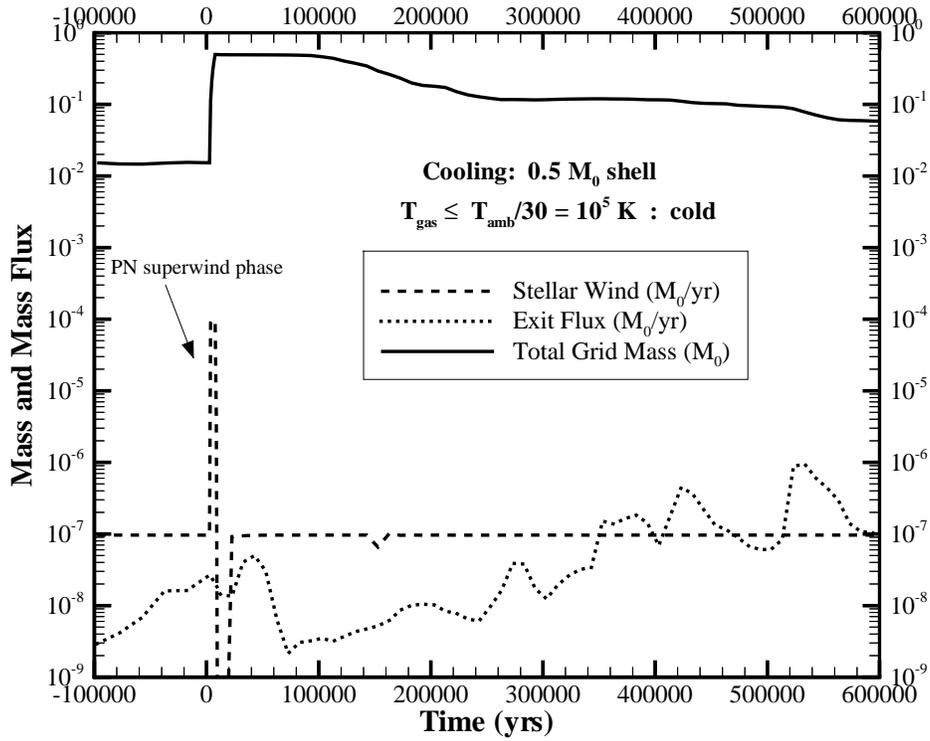}
\caption[Mass fluxes for colder $< T_{amb}/30$ K gas with cooling]%
        {Mass fluxes for colder $< T_{amb}/30$ K gas with cooling.
        The total mass and exiting mass flux curves suggest that the radiative
        cooling has managed to cool the wake, and in this case
        any PN material in the slower wake region appears to have been trapped
        there through cooling induced accretion.}
\label{fig:c0.5pncoldmass}
\end{figure}

In addition to the highly extended and detached structures, a significant amount
of cool gas remains in the wake region directly behind the star.  Momentum
transfer and mixing in this region is slower, so radiative losses more easily
balance the energy input, leading to a larger fraction of cool gas (and a longer
time to flow off the grid).  
The warm and cold gas mass flux data are shown in 
Figures~\ref{fig:c0.5pnwarmmass} and \ref{fig:c0.5pncoldmass}, respectively.
The time-lag between the PN event and the increase in exiting mass flux is
$1 \times 10^5$ years shorter than in the adiabatic case due to the increased rate
of momentum transfer.  The two large spikes in the 
exit flux curve at $2.1 \times 10^5$ years and $2.5 \times 10^5$ years are 
the two large warm/cool features close to the edge of the grid in the $2 \times 10^5$ year 
contour map (Figure~\ref{fig:cpn100.200}).  The total mass and exit flux curves
suggest that all of the warm gas ($< 7.5 \times 10^5$ K and $> 10^5$ K) has 
left the grid by $3.5 \times 10^5$ years.  

The amount of cool material is changing with time, so we estimate of the relative cooled
mass by integrating under the warm and cold exiting mass flux curves in two
intervals from $2 \times 10^5$ years to $3.5 \times 10^5$ years, and from 
$3.5 \times 10^5$ years to $6 \times 10^5$ years.  Since the first interval contains the
exiting large structures, the bulk of the gas in this interval is 
warm.  The second interval shows a growing cold wake, with the majority
of the gas exiting in this interval being cold ($< 10^5$ K).  The values
that we obtain confirm this:  the percentage of cold gas leaving the grid 
in the first interval is $\approx 3\%$; the percentage for the second interval is
$\approx 80\%$.  The cold fraction for the full $2 \times 10^5$ to $6 \times 10^5$ year 
interval, which covers the bulk of the time when PN material is leaving the 
grid, is $\approx 30\%$.  These values as well as the total grid mass data
suggest that although the warmer PN material quickly exits the grid, radiative
cooling is able to keep a significant fraction of the PN material in the slower 
wake region and cold.  This is twice the amount of cool gas that exits the grid
in the continuous mass loss case (Paper I).

\subsection{Reduced Planetary Nebula Masses}
\label{ssec-0.1PN}

Our $0.5 M_{\odot}$ planetary nebula simulation has roughly twice the mass
that is often found in a PN shell.  Although we do not expect
this difference in shell mass to effect the general nature of the PN
evolution, we have run another set of simulations where the superwind phase 
lasts only $10^3$ years, resulting in a $0.1 M_{\odot}$ shell.  These two
mass loss events bracket the typical PN shell masses observed.

\subsubsection{Simulation Without Radiative Losses}
\label{sssec-0.1PNnc}

\begin{figure}
\plotone{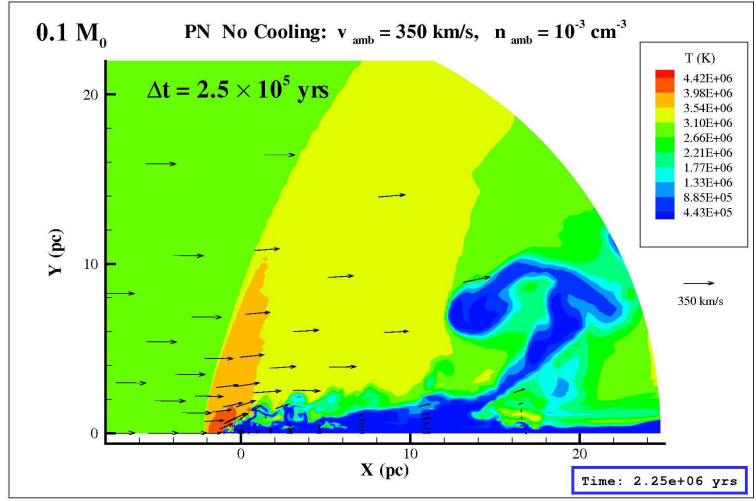}
\caption[The bulk of the smaller PN shell just prior to exiting the grid]%
        {The bulk of the smaller PN shell just prior to exiting the grid, at
        $2.5 \times 10^5$ years after the initial superwind phase.  The
        smaller mass nebula evolution is remarkably similar to the larger mass
        case:  compare this temperature map to the top panel in 
        Figure~\ref{fig:ncpn300.400}.  The general character of the
        mushroom shape is the same, although the size of the structure
        is correspondingly reduced.}
\label{fig:0.1ncpn250}
\end{figure}

The evolution of this simulation is very similar to that where the PN ejecta
is $0.5 M_{\odot}$, with just a few differences.  The rate of momentum transfer
from the ambient flow to the PN ejecta is greater than in the $0.5 M_{\odot}$ 
case.  This is because the mass of the ejecta increases as $r^3$ while the area
with which it interacts with the ambient flow increases as $r^2$, so the
velocity of the ejecta increases as $M_{ejecta}^{-1/3}$.  Due to the increased
momentum transfer, most of the gas exits the grid by $3 \times 10^5$ years.
Because the evolution is so similar to the $0.5 M_{\odot}$ case,
we will only include the grid state  at $2.5 \times 10^5$ years which shows the main part
of nebula almost about to leave the grid (Figure~\ref{fig:0.1ncpn250}).  
Upon exiting the grid, the main nebula structures are accelerated up 
to $\approx 300$ km/s, close to the speed of the ambient material.
The velocity difference between these nebular features and the hot ambient
medium is less than the sound speed in either medium, so subsequent heating
of that gas will occur primarily through turbulent mixing.

\begin{figure}
\plotone{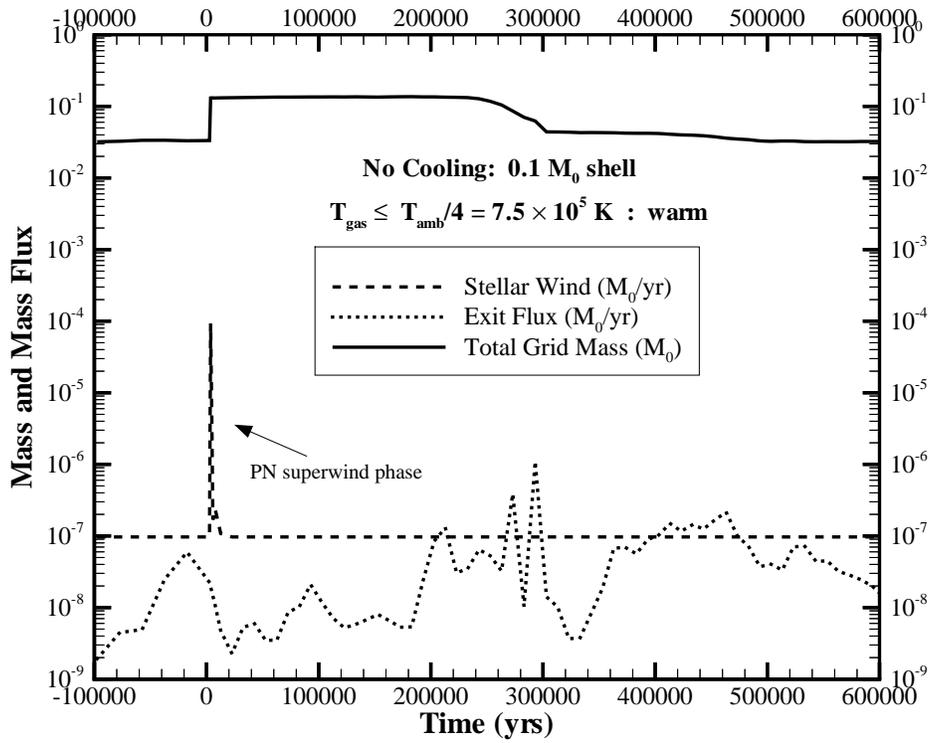}
\caption[$0.1 M_{\odot}$ mass fluxes for warm $< T_{amb}/4$ K gas]%
        {$0.1 M_{\odot}$ mass fluxes for warm $< T_{amb}/4$ K gas,
        showing a quicker advection of the PN material off the grid, and two
        main parcels of gas which leave the grid just after $2.5 \times 10^5$
        years.  These two larger parcels are shown in 
        Figure~\ref{fig:0.1ncpn250}.}
\label{fig:nc0.1pnwarmmass}
\end{figure}

\begin{figure}
\plotone{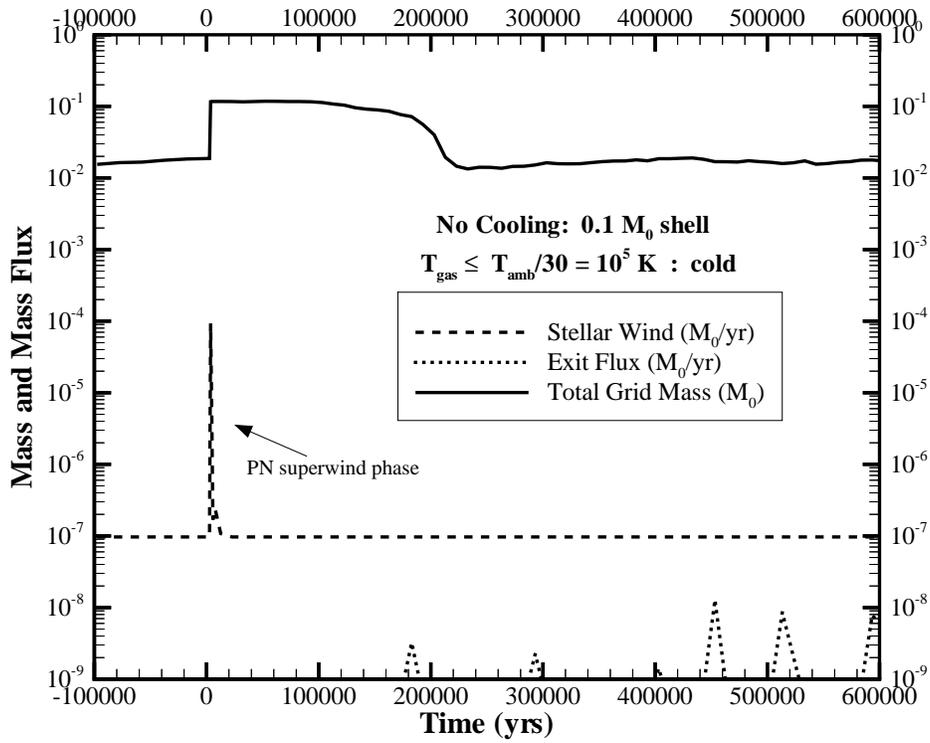}
\caption[$0.1 M_{\odot}$ mass fluxes for colder $< T_{amb}/30$ K gas]%
        {$0.1 M_{\odot}$ mass fluxes for colder $< T_{amb}/30$ K gas,
        showing the same heating effects as the larger mass nebula.}
\label{fig:nc0.1pncoldmass}
\end{figure}

An examination of the mass flux curves (Figures~\ref{fig:nc0.1pnwarmmass} and 
\ref{fig:nc0.1pncoldmass}) show that the detached PN ejecta feature exits the
grid at $2-3 \times 10^5$ years while the component in the narrower wake exits at
$3.5-5.5 \times 10^5$ years.  Both components contain warm gas but neither contain 
cool gas, which is similar to the higher mass ejecta case.

\subsubsection{Simulation With Radiative Losses}
\label{sssec-0.1PNc}

\begin{figure}
\epsscale{0.85}
\plotone{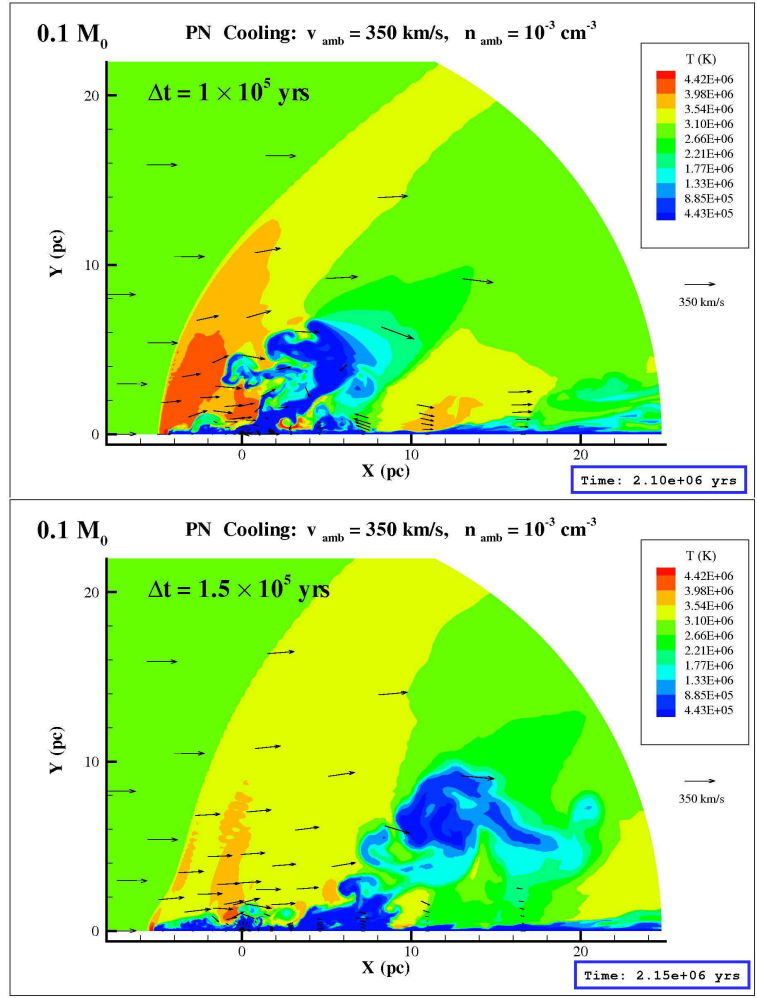}
\caption[Typical cooled instabilities and disruption of the PN]%
        {Typical cooled instabilities and disruption of the PN.  At
        $10^5$ years, the PN is in process of breaking apart, and by
        $1.5 \times 10^5$ years a large blob has broken off completely.
        This blob is quickly moving off the grid, and will be gone just after
        $2 \times 10^5$ years.}
\label{fig:0.1cpn100.150}
\end{figure}

As with the $0.5 M_{\odot}$ case, the instabilities are much stronger and more 
numerous than in the simulation without radiative losses, but the
cooling compresses many of these fingers and the wake so that they do not
heat up to the ambient temperature.  Figure~\ref{fig:0.1cpn100.150} shows
maps of the cooling run at $1 \times 10^5$ years and $1.5 \times 10^5$ years.  The top panel
shows the typical unstable nature of the ejecta with cooling included, 
and just $0.5 \times 10^5$ years later the bulk of the PN has broken off 
completely.  It exits the grid at $\approx 250$ km/s, less than in the case without
radiative losses and this is probably due to the smaller cross section that
results from the additional compression of the cooler gas.  
A large component of the nebula ejecta remains 
in the wake, remaining cooler and moving more slowly at the same timestep.

\begin{figure}
\plotone{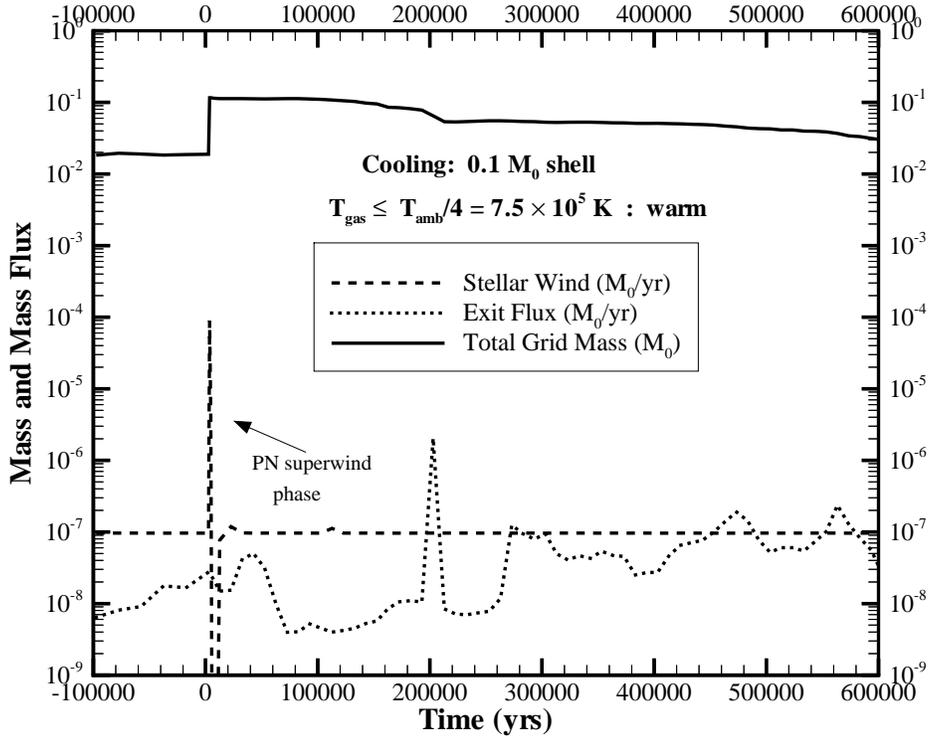}
\caption[$0.1 M_{\odot}$ mass fluxes for warm $< T_{amb}/4$ K gas with cooling]%
        {$0.1 M_{\odot}$ mass fluxes for warm $< T_{amb}/4$ K gas in the simulations with
        radiative cooling.  The majority of warmer gas that leaves the
        grid does so in the large broken-off parcel of gas that shows up as a
        sharp peak in the exiting mass flux curve just after $2 \times 10^5$ years.
        The cooling has also hastened the departure of the warm gas compared to
        the adiabatic case.}
\label{fig:c0.1pnwarmmass}
\end{figure}

\begin{figure}
\plotone{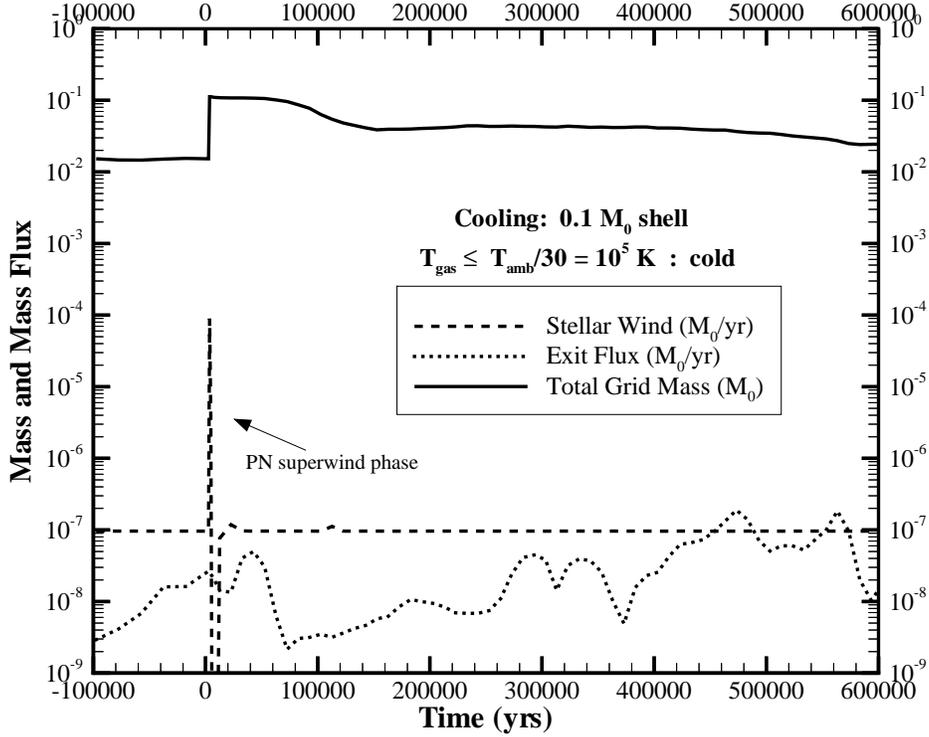}
\caption[$0.1 M_{\odot}$ mass fluxes for colder $< T_{amb}/30$ K gas with cooling]%
        {$0.1 M_{\odot}$ mass fluxes for colder $< T_{amb}/30$ K gas in the simulations with
        radiative cooling.  With the exception of the large warm parcel that quickly
        leaves the grid, remaining PN material flows along the narrow wake and
        continue to cool until it leaves the grid.  The exiting mass flux
        curve is almost the same as in the warmer case, which indicates that
        nearly all wake material leaving the grid is cold.  The coldest ejecta is 
        heated in the first $10^5$ years, which is comparable to the radiative cooling time.}
\label{fig:c0.1pncoldmass}
\end{figure}

The warm and cold mass flux curves (Figures~\ref{fig:c0.1pnwarmmass}
and \ref{fig:c0.1pncoldmass}) confirm that the cooled gas moved off the 
grid more quickly than in the non-cooling case and the ejecta is broken up 
into smaller parcels.  
The sharp peak around $2 \times 10^5$ years in the exiting mass flux curve shows 
the exit of the large component that had broken off (Figure~\ref{fig:0.1cpn100.150}).  
A comparison of the cold and warm exit mass flux curves shows that this piece contains 
only warmer gas as it exits this simulation.  However, the velocity difference
between this gas parcel and the ambient medium is about $100$ km/s, so most
of the energy and momentum has been transferred to this gas.  If further momentum
transfer were to erase the velocity difference with the ambient medium, the 
additional thermal heating of the cloud would only raise the temperature modestly.
This would still be at least a factor of three cooler than its surroundings
and radiative cooling would reduce the temperature, making this a cold cloud.
The remainder of the PN material lies in the less rapidly moving wake, which continues 
to cool radiatively as it moves downstream.  The fraction of cold gas leaving the 
grid between $2 \times 10^5$ and $6 \times 10^5$ years is about $\approx 40\%$.  
However, excluding the departure event at $2 \times 10^5$ years, $\approx 75\%$ of 
all the gas that leaves the grid has a temperature below $10^5$ K.  

\section{Discussion and Conclusions}

We simulated the interaction between the mass loss from stars during their
planetary nebula stage and the hot medium of an early-type galaxy. \ It is
necessary to model this common interaction if one is to understand the
evolution of hot and cold gas in these galaxies. \ About half the mass shed
by old stars occurs continuously, although at varying mass loss rates,
during the ascent up the giant and asymptotic giant branch, which we modeled
in Paper I. \ The other half of the mass loss occurs as a relatively rapid
event, the birth of a planetary nebula, and that stage is modeled here. \ We
simulated the ejection of the planetary nebula shell with and without
radiative losses in order to separate the physical events related to each. \
For a summary, we will focus on the simulations with radiative losses.

For most previous and the current simulations, the flow past the star is
supersonic, so a bow shock develops, which shocks both the fast ambient flow
and the stellar mass loss. \ For the previous continuous mass-loss
simulations, the mass loss expanded until the momentum of the ejecta
approached that of the fast but dilute ambient medium. \ The momentum
transfer of the shock caused stellar mass loss to flow downstream, creating
a wake, which is subject to Kelvin-Helmholtz and Rayleigh-Taylor
instabilities. Shocks propagate into  the dense extensions created by these instabilities, producing sigificant heating. 

In the
planetary nebula simulations, the ejecta expands to about 3 pc before reaching a comparable value to the momentum density of the fast flow.
\ Moving around this expanded ejecta is the flow of the post-shock hot gas,
which excites instabilities of several wavelengths, but the dominant mode is
one that is similar to the size of the expanded ejecta. \ About half of the
ejecta becomes extended vertically relative to the initial direction of the
fast flow. \ This increases the cross section for momentum transfer and that
cooler material is accelerated downstream and heated through shocks into the
originally cool ejecta. \ By the time the detached ejecta flows through the
boundary of the computational grid (25 pc from the star), it is moving
between 50-70\% of the velocity of the background flow. \ However, the
temperature of the gas is less than 25\% that of the pre-shock ambient gas.
\ The additional heating that would occur by having the cloud reach the
ambient hot gas velocity is insufficient to bring it to the
temperature of the ambient dilute medium. \ This component of the ejecta will radiatively cool, provided that the hotter ambient
medium does not mix in at a rate that exceeds the radiative cooling rate. \
Our simulation cannot address this issue.

Approximately half of the planetary nebula ejecta flows back behind the star
as a cool wake that moves at about half the speed of the detached ejecta. \
The fast flow past the wake excites fluid instabilities that lead to shocks
being driven into the gas as well as thermal mixing between the wake and the
surrounding ambient medium. \ The rate of heating associated with these
processes is less than the radiative cooling rate, so the most of the wake
material ($> 80\%$) remains cool ($< 10^{5}$ K). \
At the exit boundary, 30-40\% of the total mass of the ejecta is still cool,
which is at least double the amount from the comparable continuous mass loss
case of Paper I (15\%). \ If there is additional radiative cooling of this
ejecta beyond the computational boundary, as argued above, most of the mass
of the ejecta from a planetary nebula may remain cool. \ These results
suggest that continuous stellar mass loss, during the giant stages, is more
likely to become the hot ambient interstellar medium than the ejecta from
planetary nebula, which is more likely to remain cool.

The amount of remaining cool ejecta results from a balance between radiative
losses and fluid heating processes. \ The radiative loss rate depends upon
the metallicity, so higher metallicity gas is likely to remain cool. \ This
code does not have the ability to give one component of the fluid a
different metallicity (the ejecta), while keeping the hot ambient flow
unchanged. \ Therefore, we could not carry out a simulation to investigate
this point. \ From our previous simulations (Paper I), we found that the
results did not always change as anticipated or by the magnitude that one
might infer from simple estimations. \ From these experiences, a careful
simulation will be needed to address the metallicity issue.

One of the suggestions presented in Paper I is that at least some of the
optical emission line gas (at $\sim 10^{4}$ K) that is seen in early-type
galaxies is the cool mass loss from evolving stars. \ Gas emitting at higher
temperatures, such as OVI, is seen in observations obtained by the FUSE
satellite \citep{breg05}. \ The ion OVI exists primarily in the $2-5\times 10^{5}$ K
temperature range, and this is similar to the temperatures found in the wake
and the detached ejecta (or the gas has passed through this range in cooling). \ The
observed OVI line fluxes has been interpreted as indicative of gas cooling from
the ambient temperature though the OVI temperature range (and eventually to $%
10^{4}$ K or lower \citep{edgar86}; this is a cooling flow model). \ Our calculations
suggest that an alternative explanation is possible: \ the collective
transient heating of stellar ejecta to the OVI temperature range. \
Unfortunately, non-equilibrium ionization effects are important in this
temperature range, so to calculate the OVI\ line fluxes accurately, one will
need to follow the ionization properties of every fluid element, which is
well-beyond the capabilities of our calculations.

While our calculations indicate certain types of behavior for the mass loss
from stars, it will require a more ambitious set of calculations to properly
determine the fractional mass of cool gas and the emission line strengths. \
Of central importance has been the strength and evolution of fluid
instabilities, but in our calculation, the fluid is treated in two spatial
dimension, with symmetry in the third (angular) coordinate. \ This may
change the growth rate of instabilities, and consequently, the evolution of
the stellar ejecta. \ Another improvement would be simulations that could
follow the late time evolution of the stellar ejecta (beyond our grid
boundaries), and even the evolution of the material as it begins to fall
inward in the galaxy. \ An important addition would be the ability to
identify the stellar mass loss (as opposed to the ambient material) and for
the ejecta to have a different metallicity than the ambient gas. \ Lastly, a
larger grid of simulations is warranted, such as simulations in which the
ambient pressure is significantly larger, such as in the central few hundred
pc of a galaxy.

\acknowledgements
JRP would like to give special thanks to Philip Hughes and Hal Marshall for their
valuable advice and comments during code development, simulations, and on the
content of my PhD Thesis.  Also, thanks are due to Gus Evrard and Douglas Richstone
for serving on JRP's thesis committee and for providing timely support and advice.
Financial support is gratefully acknowledged and was provided to JRP through
a Department of Energy Computational Science Graduate Fellowship and to JNB through
a Long Term Space Astrophysics Grant from NASA.

\clearpage

\clearpage

\begin{figure}
\epsscale{0.85}
\plotone{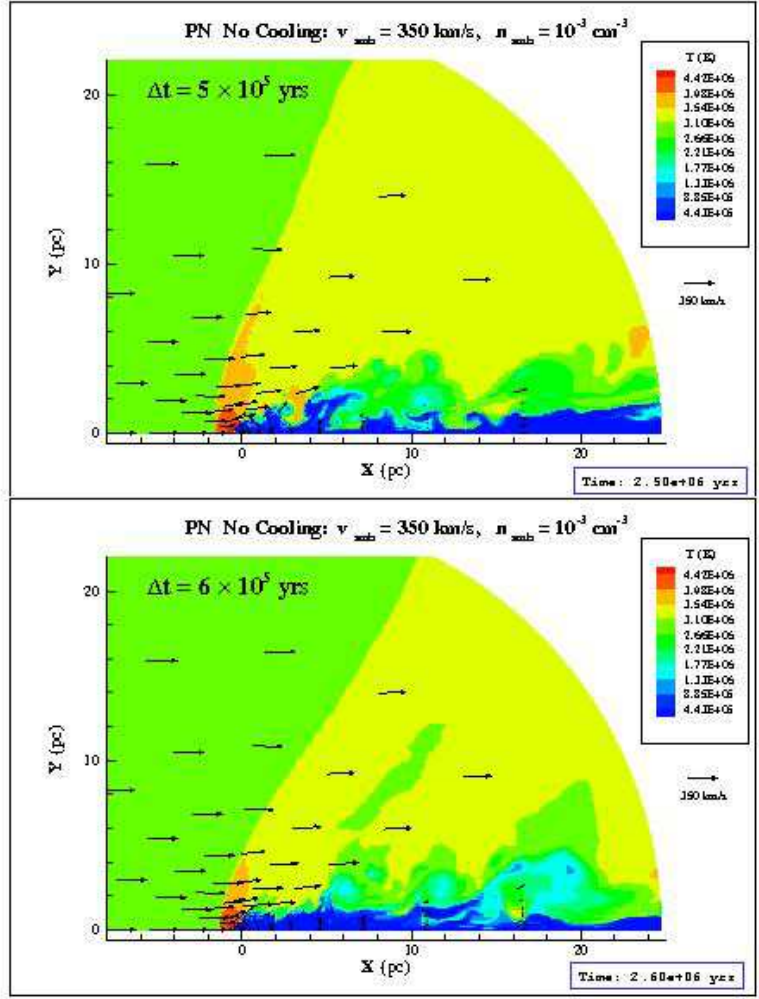}
\caption[The vast majority of the PN has exited the grid]%
        {The majority of the PN has exited the grid, although some material
        may remain in the slow wake.  By $6 \times 10^5$ years (bottom), the
        flow has nearly returned to its pre-PN state.}
\label{fig:ncpn500.600}
\end{figure}

\clearpage

\clearpage

\clearpage

\clearpage

\begin{figure}
\epsscale{0.85}
\plotone{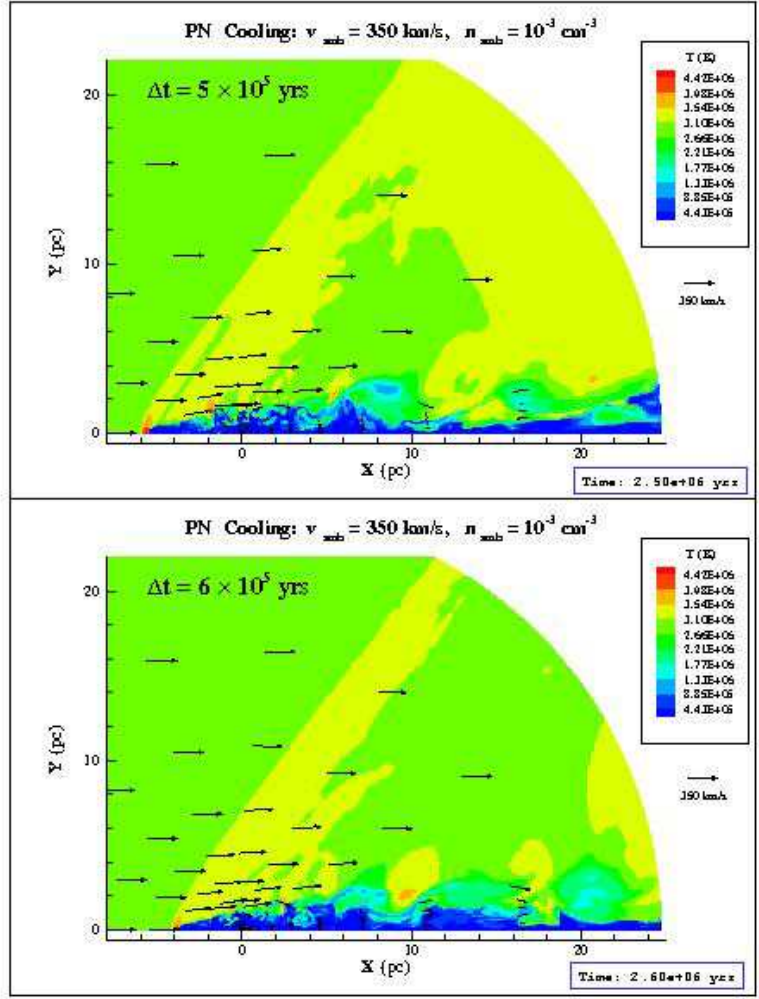}
\caption[The flow has returned to the pre-PN state]%
        {The flow has returned to the pre-PN state, although more material
        may remain in the cooled slow wake.}
\label{fig:cpn500.600}
\end{figure}

\clearpage

\end{document}